# Two-color flat-top solitons in microresonator-based optical parametric oscillators

Valery E. Lobanov

*Russian Quantum Center, Skolkovo 143026, Russia*

**ABSTRACT**

We studied numerically the generation of the two-color flat-top solitonic pulses, platicons, in the microresonator based doubly resonant optical parametric oscillator. We revealed that if the signs of the group velocity dispersion (GVD) coefficients at interacting harmonics are opposite, platicon excitation is possible via pump amplitude modulation or controllable mode interaction approach. Upon pump frequency scan generation of platicons was observed at positive pump frequency detunings for the normal GVD at pump frequency and at negative detunings in the opposite case. Interestingly, we found the effect of the transformation of the flat-top platicon profile at half-frequency into the bell-shaped bright soliton profile upon frequency scan. For platicon excitation one needs simultaneous accurate matching of the microresonator free spectral ranges at interacting harmonics and resonant eigenfrequencies. Excitation conditions and platicon generation domains were found for the different generation methods, and properties of the generated platicons were studied for various combinations of the medium parameters.

## 1. INTRODUCTION

Over the past two decades, optical frequency combs have established themselves as a unique and indispensable tool both for scientific research and important technological applications [1, 2]. Microresonator-based Kerr frequency combs (or microcombs) are of particular interest due to the high compactness and energy efficiency of the comb generator [3-8]. Since their discovery in 2007 [3] they became an object of the intensive investigations and were demonstrated in high-Q microresonators made of different materials (crystalline fluorides, diamond, quartz, silicon, silicon nitride, etc.) and of different geometries, bulk and on-chip. Further, it was shown the possibility of the generation of the coherent microresonator-based frequency combs in the form of the dissipative Kerr solitons [9,10]. Such structures were successfully used in various important up-to-date applications, such as high-precision metrology [11,12], high-resolution spectroscopy [13,14], astrophysics [15,16], and high-volume telecommunication systems [17]. Recently, it was shown that generation of the optical frequency combs is also possible in materials with quadratic nonlinearity, such as $LiNbO_3$ or $LiTaO_3$ [18-27]. This fact aroused great interest among researchers, since generation of optical frequency combs due to the quadratic nonlinearity may be realized at reduced pump powers because of the high level of the nonlinear response and in the spectral diapasons inaccessible for the conventional Kerr frequency combs. Also, it was shown that quadratic nonlinearity may support different types of the dissipative solitons (bright, dark and quasi-solitons) in optical microresonators upon second harmonic generation process [28, 29] or via downconversion process (degenerate optical parametric oscillations) [30-34]. However, in a greater part of the previous studies the main attention was paid to the existence domains and properties of the localized states but not to their generation. In our work we studied numerically the generation of the particular type of the solitonic structures, two-color flat-top solitonic pulses, platicons, in the microresonator-based optical parametric oscillator via frequency scan that is the conventional method of the dissipative Kerr solitons generation in experiments [9]. Dark solitons [35-40] and platicons [41-46] are well-studied in Kerr microresonators and it was shown that in terms of the pump-to-comb conversion efficiency the generation of platicons may be significantly more efficient than the generation of bright solitons [47,48] that is very promising for the coherent optical communications [49, 50]. In [51] platicon generation in the quadratically nonlinear microresonators was studied for the second harmonic generation process. In our work we demonstrated numerically that using the methods developed for the platicon generation in Kerr microresonators platicon generation can be also realized for the downconversion process (degenerate optical parametric oscillations). We revealed that platicon excitation is possible if the signs of the group velocity dispersion (GVD) coefficients at interacting harmonics are opposite. Upon frequency scan the generation of platicons was observed at positive pump frequency detunings for the negative pump GVD coefficient (and positive subharmonic GVD coefficient) and at negative detunings in the opposite case. Interestingly, in some cases we observed the transformation of the flat-top platicon profile at the half-frequency into the bell-shaped bright soliton profile upon frequency scan. For both methods, for the efficient platicon excitation one needs simultaneous accurate matching of both microresonator free spectral ranges at interacting harmonics and resonant eigenfrequencies. Excitation conditions and platicon generation domains were found for different generation methods, and properties of generated platicons were studied for the different combinations of medium parameters.

## 2. MODEL

For numerical analysis we used the system of the two coupled equations for the fundamental wave (FW) and second harmonic (SH) fields which, in normalized form, may be written as

$$\begin{cases} \dfrac{\partial u}{\partial \tau} + d \dfrac{\partial u}{\partial \varphi} = i\dfrac{1}{2}b_{21}\dfrac{\partial^2 u}{\partial \varphi^2} + ivu^* - (\dfrac{\kappa_1}{\kappa_2} + i\alpha_1)u, \\ \dfrac{\partial v}{\partial \tau} = i\dfrac{1}{2}b_{22}\dfrac{\partial^2 v}{\partial \varphi^2} + iu^2 - (1 + i\alpha_2)v + f, \end{cases} \quad (1)$$

where $u$ and $v$ are the normalized slowly varying envelopes of the FW and SH fields, respectively. In comparison with SHG system described in [50], the pump term is in the second equation. Here $\tau = \kappa_2 t / 2$ denotes the normalized time, $\kappa_{1,2} = \omega_{01,02} / Q_{1,2}$ denotes the FW (SH) cavity decay rate, $\omega_{02}$ is the microresonator eigenfrequency, closest to the pump frequency $\omega_p$, $\omega_{01}$ is the microresonator eigenfrequency, nearest to the half pump frequency, $Q_{1,2}$ is the total quality factor at the fundamental frequency/second harmonic, $\varphi \in [-\pi;\pi]$ is an azimuthal angle in a coordinate system rotating with the angular frequency equal to the microresonator free spectral range (FSR) $D_{12}$ at the pumped mode, $d = 2(D_{11} - D_{12})/\kappa_2$ is the normalized difference between FSRs at the fundamental and second harmonic frequencies (temporal walk-off term), $\alpha_2 = 2(\omega_{02} - \omega_p)/\kappa_2$ is the normalized pump frequency detuning, $\alpha_1 = 2(\omega_{01} - 0.5\omega_p)/\kappa_2 = 2(\omega_{01} - 0.5\omega_{02})/\kappa_2 + (\omega_{02} - \omega_p)/\kappa_2 = 0.5(\alpha_2 - \delta)$, where $\delta = 2(\omega_{02} - 2\omega_{01})/\kappa_2$ is the normalized offset between the SH resonant frequency $\omega_{02}$ and doubled frequency of the fundamental resonance $\omega_{01}$. $f$ stands for the dimensionless pump amplitude. $b_{21}$ and $b_{22}$ are the normalized GVD coefficients at the fundamental and second harmonic frequencies, respectively. Positive/negative GVD coefficients correspond here to the anomalous/normal GVD, respectively.

We studied numerically nonlinear processes arising upon pump frequency scan across the fundamental frequency resonance ($\omega_p = \omega_p(0) - \Omega t$) with a noise-like input for different combinations of the GVD coefficients $b_{21,22}$. This method is widely used in experiments for the dissipative Kerr solitons generation [9,10]. Taking into account that an important parameter of Eq. 1 is the difference between the pump frequency and microresonator resonant frequency, and assuming microresonator eigenfrequencies to be constant, in numerical simulations it is convenient to introduce linear-in-time variation of the pump frequency detuning $\alpha_2$ ($\alpha_2 = \alpha_2(0) + \beta\tau$, $\beta = \dfrac{4}{\kappa_2^2}\Omega$) and the consequent variation of $\alpha_1 = 0.5(\alpha_2 - \delta)$.

Eq. 1 was solved numerically using a standard split-step Fourier routine with 1024 points in the azimuthal direction. We also checked that results do not change with increase of the number of the transverse points. For analysis, we calculated the dependencies $U_{1,2}(\alpha_2)$, where $U_1 = \int_{-\pi}^{\pi} |u|^2 d\varphi$ and $U_2 = \int_{-\pi}^{\pi} |v|^2 d\varphi$ are FW and SH intracavity powers, and studied the field distribution evolution upon frequency scan.

We set $d = 0$, $f = 15$, $\kappa_2/\kappa_1 = 1$. In order to guarantee the generation of the steady-state structures and to discriminate them from the transient distributions, the normalized frequency scan velocity $\beta$ was chosen rather small ($\beta = 0.002$) and it was checked that dynamics of the considered processes does not change if $\beta$ decreases further. We also considered the case of the ideal matching of the resonant frequencies ($\delta = 0$), thus $\alpha_2 = 2\alpha_1$. Simultaneous matching of FSRs and resonant frequencies can be done by the correct choice of the pumped mode and fine tuning of the microresonator geometry [29,33]. Phase matching of the microresonator may be realized through periodic poling [5,52]. Note that resonant conditions for the different polarizations of the interacting waves may correspond to the different combinations of the signs of the GVD coefficients.

Similarly to the previously studied case of the SHG [50], stable high-intensity branches were observed for the case of the opposite signs of the GVD coefficients. The difference is that while for the SHG process such branches were found for both positive and negative detunings for both combinations of the GVD coefficients, for the downconversion process studied here such branches were observed at positive detunings for $b_{21} = 0.05$, $b_{22} = -0.05$ and at negative detunings for $b_{21} = -0.05$, $b_{22} = 0.05$. As it was shown in [50], this fact gives us an opportunity for generation of the two-color localized structures similar to the platicons in Kerr microresonators.

### 3. PLATICONS VIA PUMP MODULATION

First, we studied the method based on the pump amplitude modulation with the modulation frequency equal to the microresonator FSR [42,44]. To take pump modulation into account we replaced the homogeneous pump term $f$ in Eq. 1 by the modulated one $f(1+\varepsilon\cos\varphi)$, where $\varepsilon$ is the modulation depth. Studying field distribution evolution for $b_{21} = 0.05$, $b_{22} = -0.05$ upon frequency scan (see top panels in Fig. 1) one may notice that, first, at negative detunings the generation of some indented patterns may be observed. These structures look like amplitude-modulated Turing patterns (see top left panel in Fig. 2). They experience modifications upon detuning growth and approximately at $\alpha_2 = -5$ transform into wide patterns with several deep oscillations at the ends (see top right panel in Fig. 2). Then such pattern becomes unstable at $2.3 < \alpha_2 < 3.1$ (see bottom left panel in Fig. 2) and turns into smooth amplitude-modulated profile. At $\alpha_2 = 6.15$ the generation of the wide localized state, platicon, takes place.

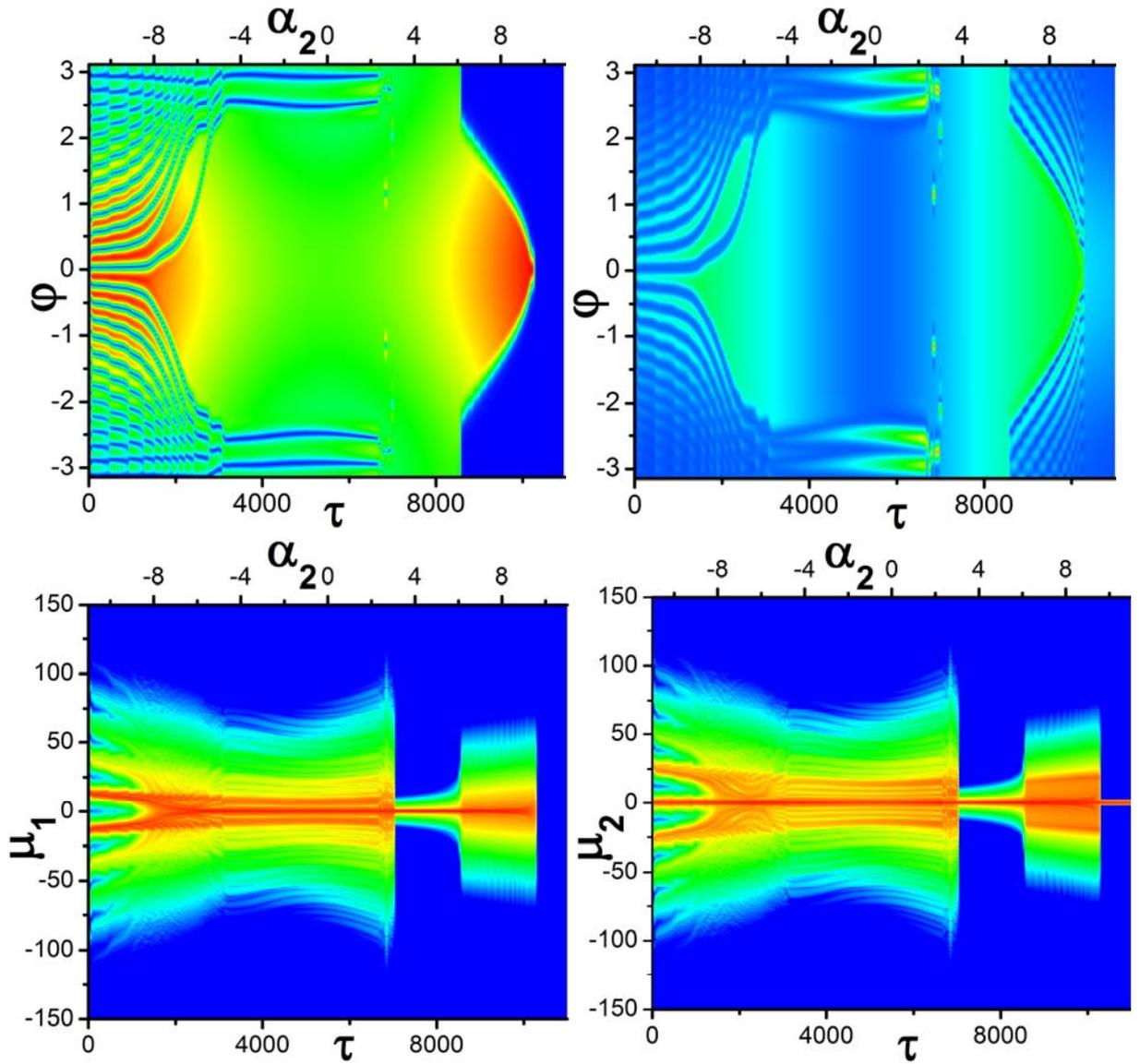

**Fig. 1.** (left panels) FW and (right panels) SH field distribution (upper line) and spectrum (bottom line, logarithmic scale) evolution upon forward pump frequency scan $\alpha_2 = -11.0 + 0.002\tau$ at $b_{21} = 0.05$, $b_{22} = -0.05$, $f = 15$, $\varepsilon = 0.4$. $\mu_{1,2}$ are mode numbers, $\mu_2 = 0$ corresponds to the pumped mode, $\mu_1 = 0$ – to the mode nearest to the half pump frequency. All quantities are plotted in dimensionless units.

In the frequency domain, one may notice a sudden significant broadening of the spectrum at $\alpha_2 > 6.15$ (see bottom panels in Fig. 1, where mode numbers $\mu_2$ are defined relative to the pumped mode corresponding to $\mu_2 = 0$ and $\mu_1$ – relative to the mode with the eigenfrequency nearest to the half pump frequency) at platicon generation. All these stages of the field distribution evolution, including formation of periodic patterns, unstable regime, platicon excitation, are also clearly visible in the dependencies $U_{1,2}(\alpha_2)$ (see Fig. 3).

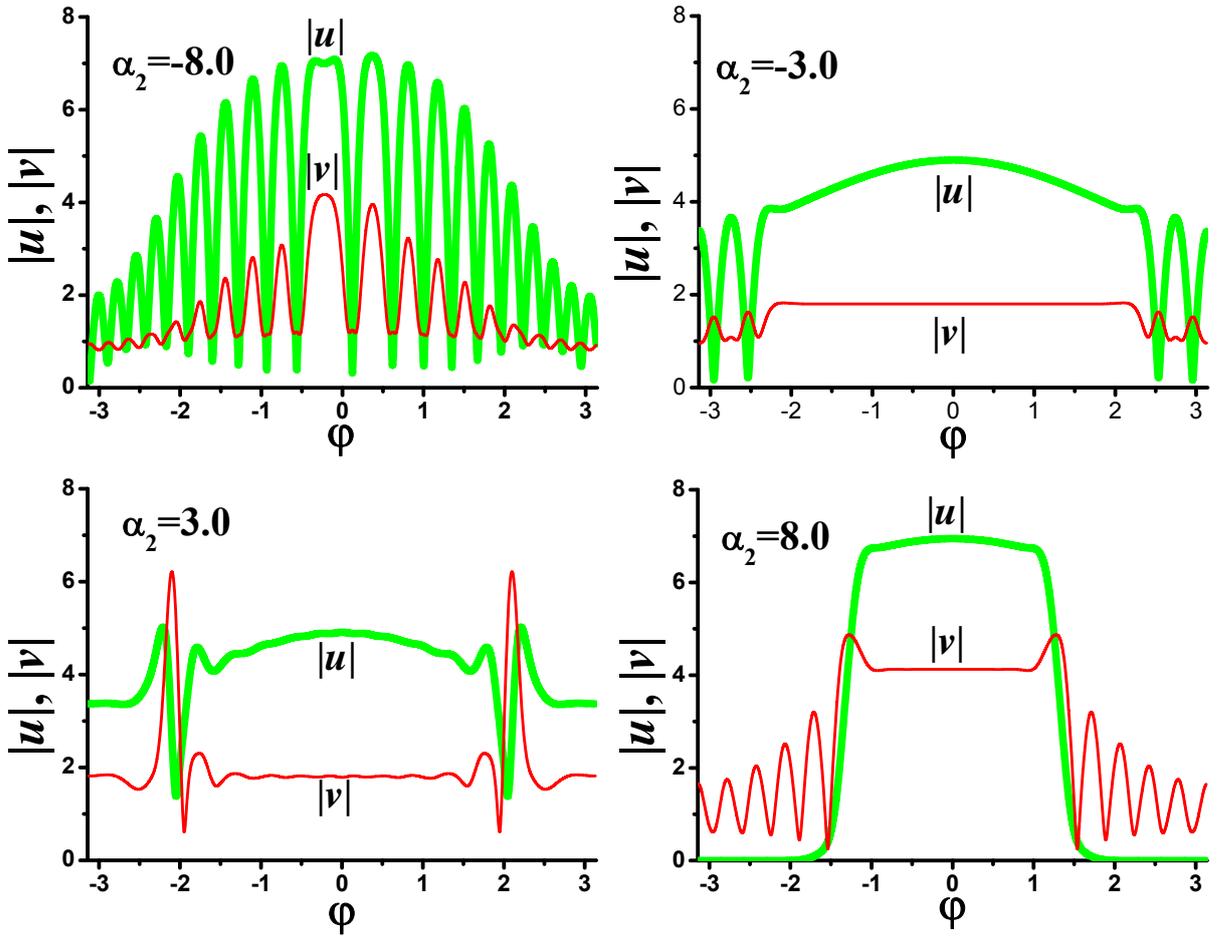

**Fig. 2.** Profiles of the FW and SH components of the generated patterns for $\alpha_2 = \pm 8.0$ and $\alpha_2 = \pm 3.0$ at $f = 15$, $\varepsilon = 0.4$ for $b_{21} = 0.05$, $b_{22} = -0.05$. All quantities are plotted in dimensionless units.

Interestingly, platicon profiles of the pumped wave (SH) are characterized by the pronounced oscillating tails, while they are absent at FW profiles (see bottom right panel in Fig. 2). Platicons become narrower with the growth of the detuning absolute value.

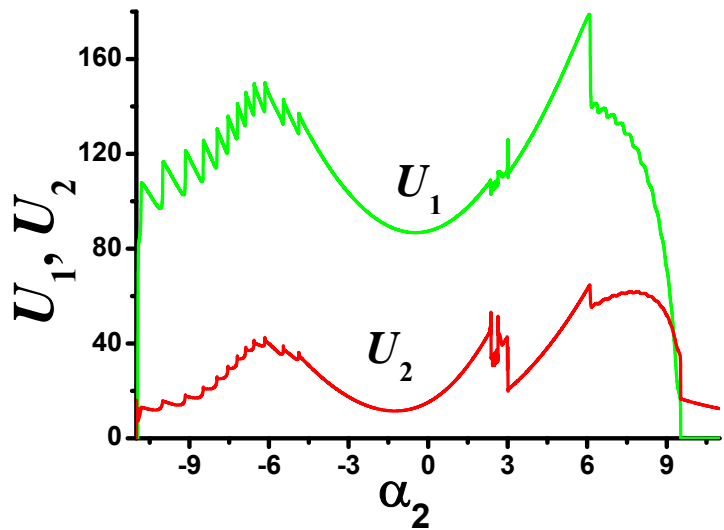

**Fig. 3.** FW and SH intracavity powers $U_1$ and $U_2$ vs. pump frequency detuning $\alpha_2$ upon forward pump frequency scan $\alpha_2 = -11.0 + 0.002\tau$ at $b_{21} = 0.05$, $b_{22} = -0.05$, $f = 15$, $\varepsilon = 0.4$. All quantities are plotted in dimensionless units.

Platicon excitation occurs if the modulation depth is larger than some critical value, depending on the pump amplitude. Generation domain becomes wider with the growth of the modulation depth (see the left panel in Fig. 4) and shifts to the larger absolute values of the detuning with the growth of the pump amplitude (see the right panel in Fig. 4).

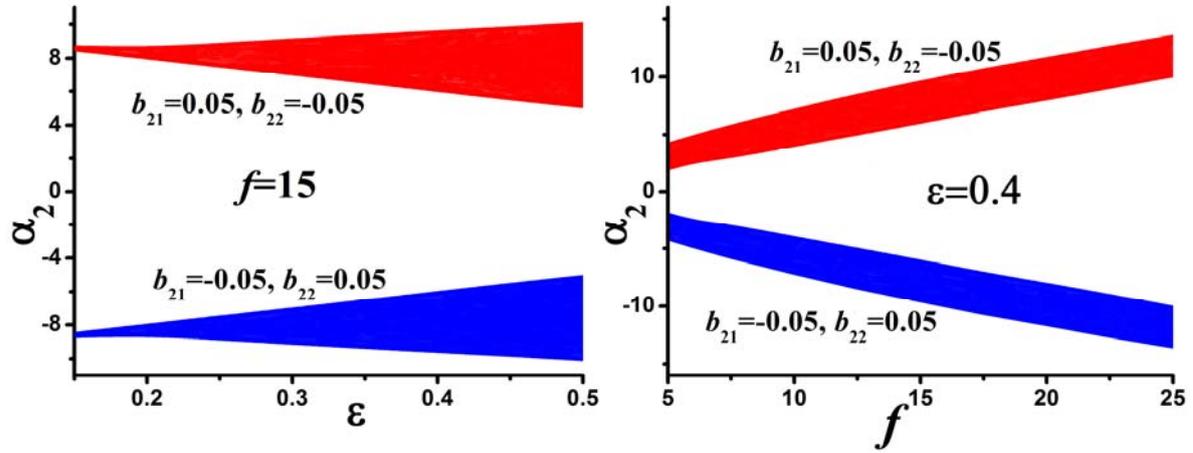

**Fig. 4.** Platicon generation domains for different values of the modulation depth and fixed pump amplitude (left panel) and for different pump amplitudes and fixed modulation depth (right panel) at $b_{21} = 0.05$, $b_{22} = -0.05$ and $b_{21} = -0.05$, $b_{22} = 0.05$. All quantities are plotted in dimensionless units.

We found that considered method is applicable for the wide range of the material and pump parameters. Platicon generation was observed for the different pump amplitudes (at least for $f = 5...25$), different absolute values of the GVD coefficients ($|b_{21}| = 0.025...0.25$). The most simulations were carried out for the same absolute values of FW and SH GVD coefficients ($|b_{21}|=|b_{22}|$). However, it was revealed that platicons can be generated if absolute values of the GVD coefficients are not equal. For example at $f = 15$, $\varepsilon = 0.4$ and $|b_{21}| = 0.05$) platicon generation was observed at $0.1 < |b_{22}/b_{21}| < 2.5$. Interestingly, when $|b_{22}/b_{21}| < 1$, we observed some kind of a transition from platicon generation to soliton generation upon frequency scan (see Fig. 5).

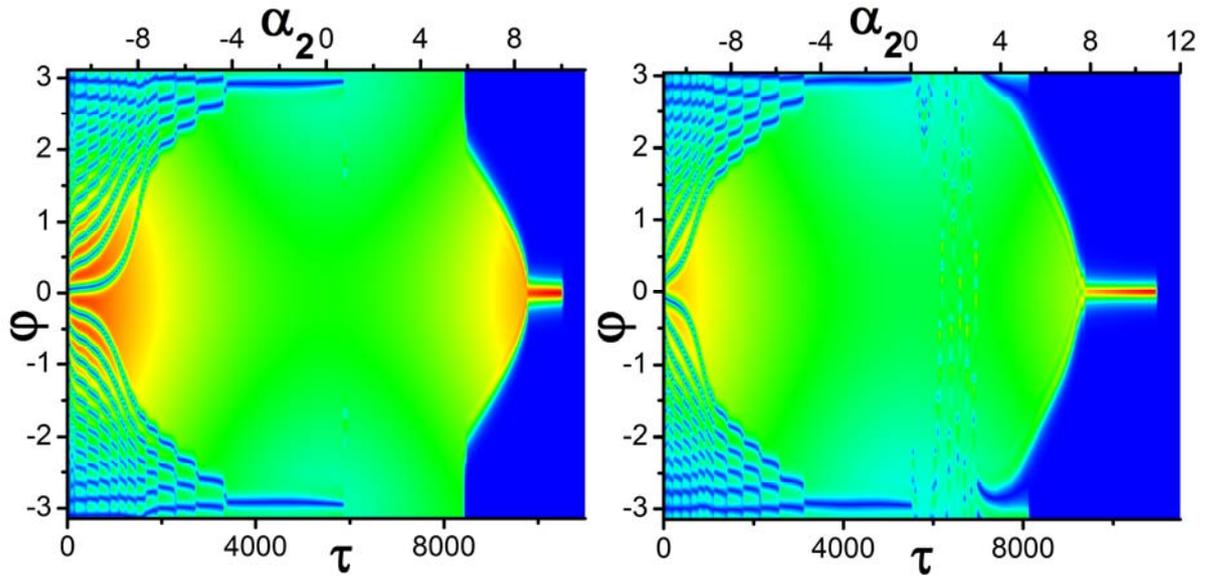

**Fig. 5.** FW field distribution evolution upon forward pump frequency scan $\alpha_2 = -11.0 + 0.002\tau$ at $b_{21} = 0.05$, $f = 15$, $\varepsilon = 0.4$ for (left panel) $b_{22} = -0.02$ and (right panel) $b_{22} = -0.01$. All quantities are plotted in dimensionless units.

Such transition was confirmed by the change of the dependence of the intracavity power on the pump frequency detuning $U_1(\alpha_2)$ and by the spectrum transformation (see Fig. 6). At $b_{22} = -0.05$ only the platicon generation occurs. For other cases shown in the left panel in Fig. 5 soliton steps are also observed. One may notice the "wings" at the spectrum profile, characteristic for platicons, at $b_{22} = -0.05$ (blue curve in the right panel in Fig. 6) and their absence at $b_{22} = -0.02$ (green curve) and $b_{22} = -0.01$ (red curve). Transitions between different types of localized states are actively studied in different nonlinear systems (see e.g. [53-57]) and, thus, observed phenomenon seems to be very intriguing and deserves more detailed research in the future. Earlier, coexistence of the dark and bright solitons in Kerr microresonators was predicted in the case of the normal GVD with the presence of the third-order dispersion [58]. Transformation of the square-like profiles into bell-shaped was also observed in microring optical parametric oscillators due to the competition of the $\chi^{(2)}$ and Kerr nonlinearities [33].

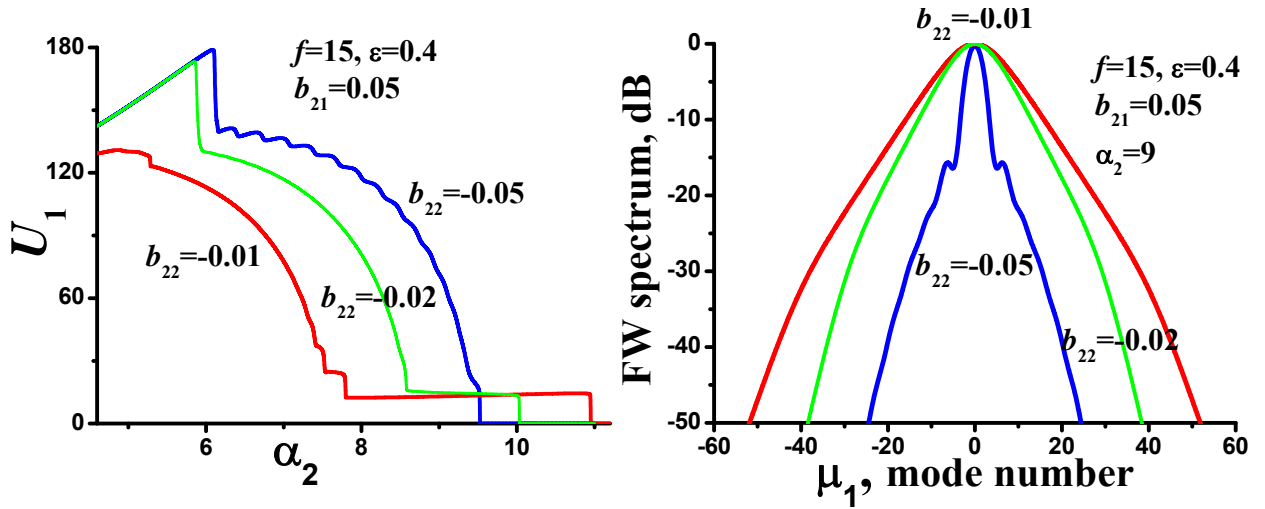

**Fig. 6.** (left) $U_1(\alpha_2)$ and (right) FW spectrum for different values of $b_{22}$ at $b_{21} = 0.05$, $f = 15$, $\varepsilon = 0.4$. $\mu_1 = 0$ corresponds to the mode nearest to the half pump frequency. All quantities are plotted in dimensionless units.

We also checked that this method is very sensitive to the matching of FSRs at the interacting harmonics. Platicon generation was observed for $|d| \leq 0.3$ at $f = 15$, $\varepsilon = 0.4$ and for $|d| < 0.36$ at $f = 15$, $\varepsilon = 0.5$. The repetition rate of the generated platicons was equal to the modulation frequency, but their profiles became asymmetric (see Fig. 7).

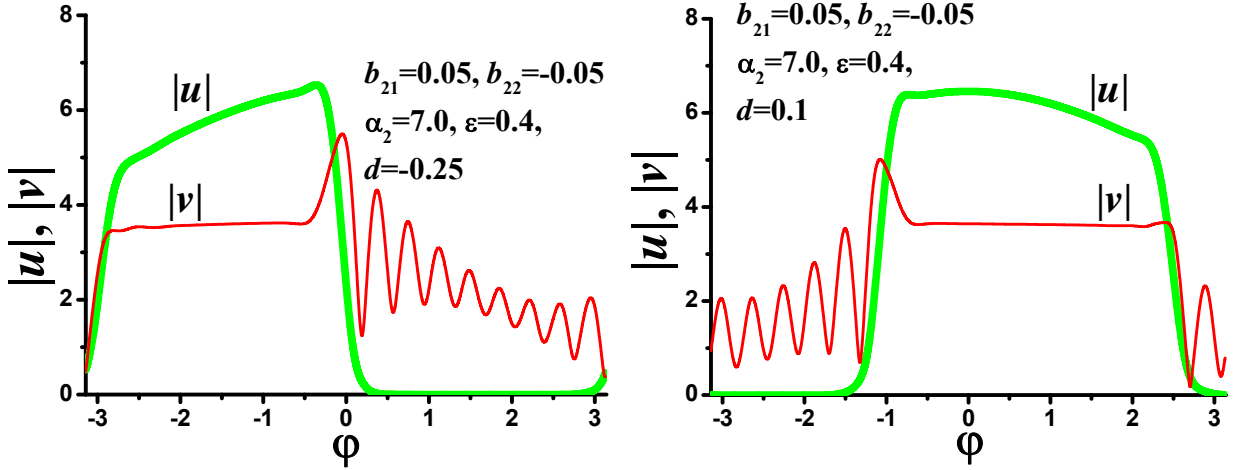

**Fig. 7.** Profiles of the platicon components for different FSRs mismatch values at $f = 15$, $\varepsilon = 0.4$, $b_{21} = 0.05$, $b_{22} = -0.05$. All quantities are plotted in dimensionless units.

The condition of the resonant frequencies matching also should be satisfied rather accurately, but admissible mismatch range is very asymmetric according to the point $\delta = 0$. At $f = 15$ and $\varepsilon = 0.4$ platicon generation was observed at $-40 < \delta < 7$ for $b_{21} = 0.05$, $b_{22} = -0.05$ and at $-7 < \delta < 40$ for $b_{21} = -0.05$, $b_{22} = 0.05$.

### 4. PLATICONS VIA CONTROLLABLE MODE INTERACTION

The second method that we studied is based on the coupling between different mode families [59-61]. Controllable mode interactions may be realized e.g. in a system of the two coupled microresonators [48, 61]. It is possible to describe such a system introducing the shift of the pumped mode [41,43,60]. For numerical analysis, we introduced the frequency shift $\Delta$ of the pumped mode eigenfrequency $\omega_{02}$: $\bar{\omega}_{02} = \omega_{02} - \Delta$. To take it into account, additional phase shift defined by $\Delta$ was applied to the central mode in the frequency domain step of the split-step Fourier routine. Thus, for the pumped mode, pump detuning $\alpha_2$ should be replaced by the effective detuning $\alpha_{20} = \alpha_2 - (2\Delta / \kappa_2)$. Platicon generation was also observed under the condition of the pumped mode shift at positive shift and mostly positive detunings at $b_{21} = 0.05$, $b_{22} = -0.05$ (see Fig. 8) and at negative shift and mostly negative detunings at $b_{21} = -0.05$, $b_{22} = 0.05$.

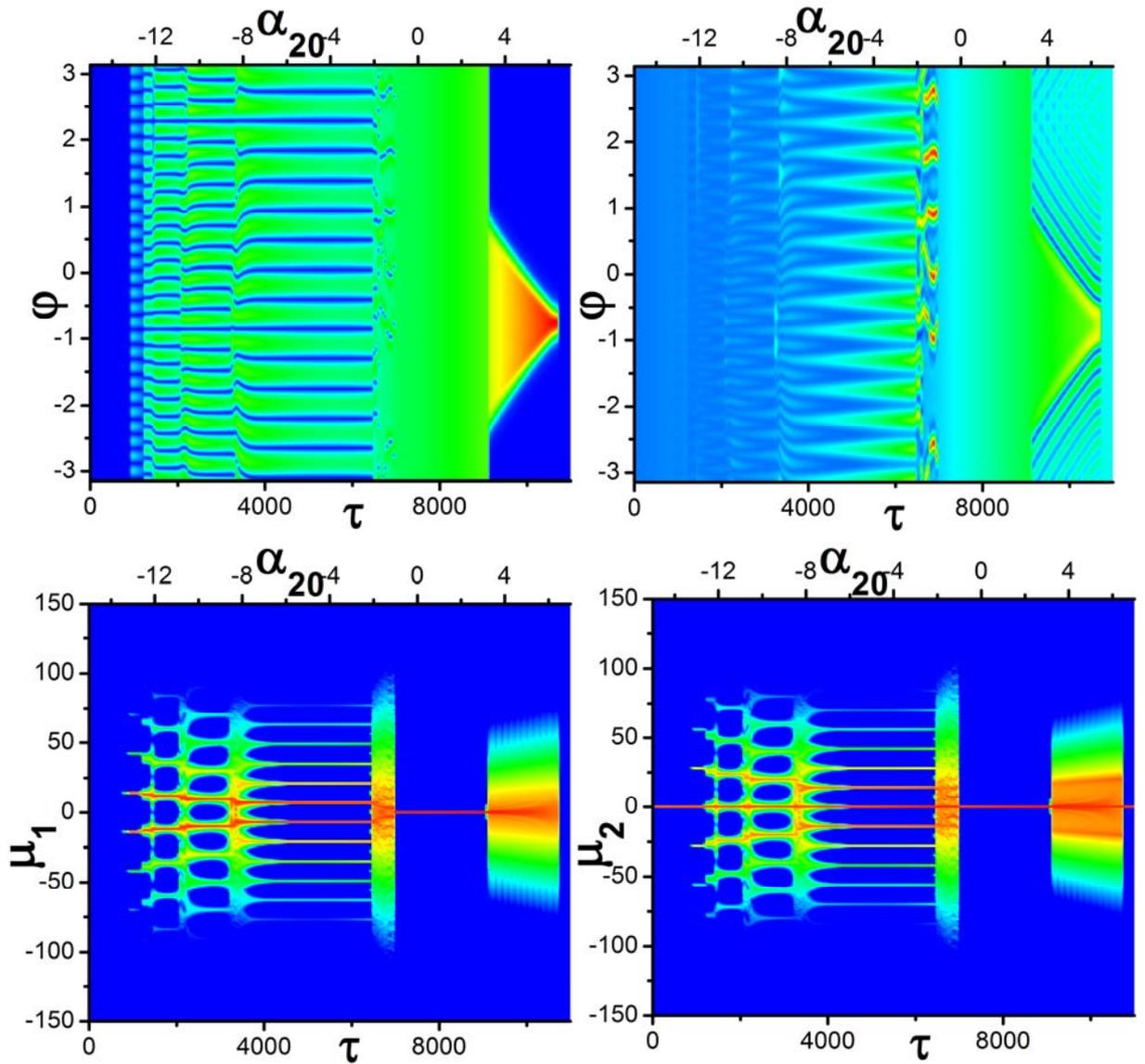

**Fig. 8.** (left panels) FW and (right panels) SH field distribution (upper line) and spectrum (bottom line, logarithmic scale) evolution upon forward pump frequency scan $\alpha_{20} = -15.0 + 0.002\tau$ at $b_{21} = 0.05$, $b_{22} = -0.05$, $f = 15$, $\Delta = 2\kappa_2$. $\mu_{1,2}$ are mode numbers, $\mu_2 = 0$ corresponds to the pumped mode, $\mu_1 = 0$ – to the mode nearest to the half pump frequency. All quantities are plotted in dimensionless units.

Besides the significant transformation of the field distribution at platicon generation (see Fig. 9), one may also notice the drastic spectrum widening (see Fig. 8, bottom line). Note that field distribution evolution upon pump frequency scan at pumped mode shift and corresponding generated patterns (Turing-like patterns shown at top panels in Fig. 9, unstable pattern at bottom left panel in Fig. 9, homogeneous solution, platicon) are quite similar to those observed at amplitude modulation (compare Figs. 2 and 9 and Figs. 3 and 10).

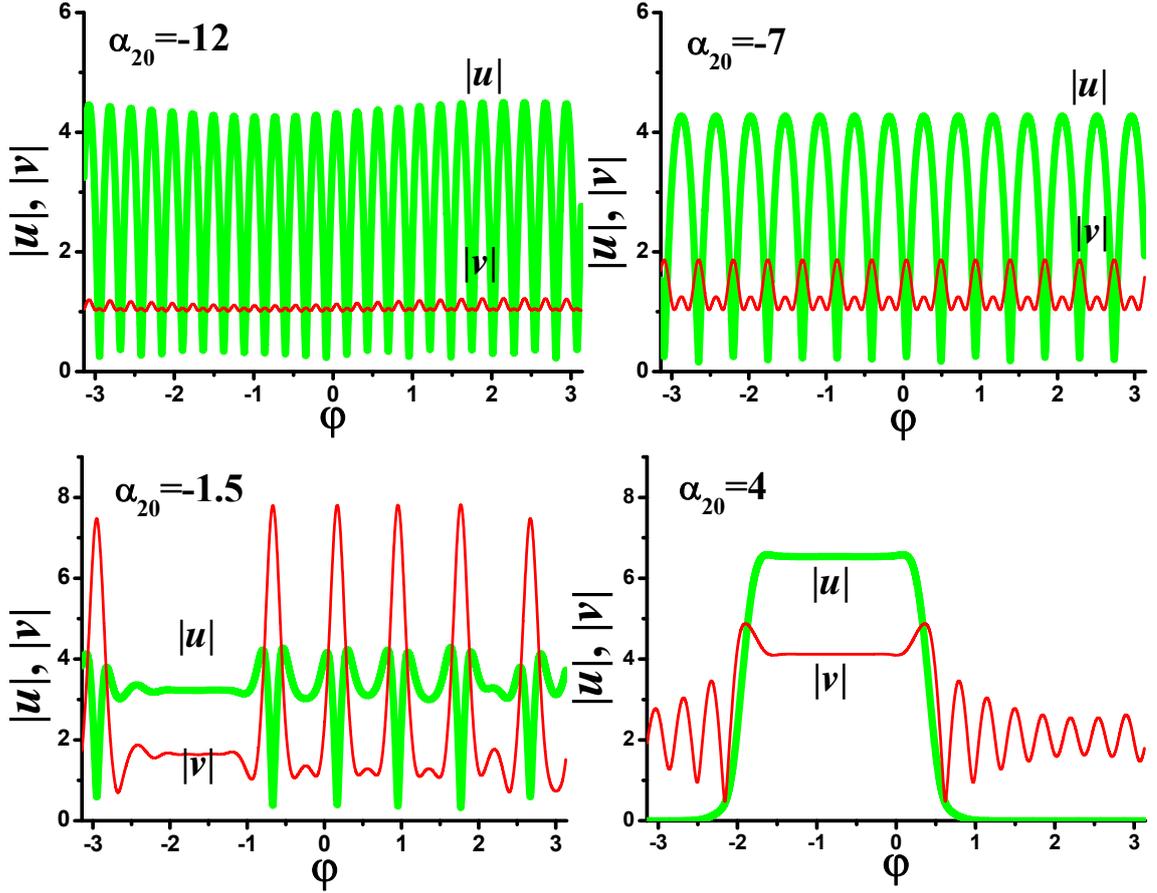

**Fig. 9.** Profiles of the FW and SH components of the generated patterns for different detuning values at $f = 15$, $\Delta = 2\kappa_2$, $b_{21} = 0.05$, $b_{22} = -0.05$. All quantities are plotted in dimensionless units.

The platicon SH component has pronounced oscillating tails, while the FW component has a smooth profile (see bottom right panel in Fig. 9).

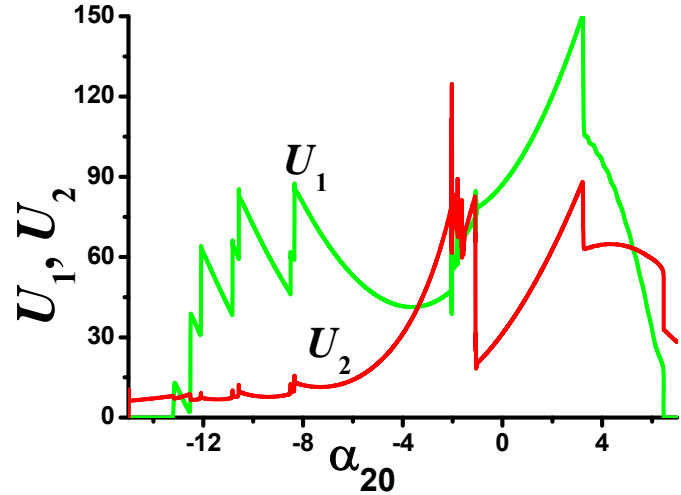

**Fig. 10.** FW and SH intracavity powers $U_1$ and $U_2$ vs. pump frequency detuning $\alpha_{20}$ upon forward frequency scan $\alpha_{20} = -15.0 + 0.002\tau$ at $b_{21} = 0.05$, $b_{22} = -0.05$, $f = 15$, $\Delta = 2\kappa_2$ All quantities are plotted in dimensionless units.

Platicon excitation was observed for the pumped mode shift with the absolute value larger than some critical value depending on the pump power (see left panel in Fig. 11). For $\Delta = 2\kappa_2$ we observed platicon generation for $f = 7.5...25$ (see right panel in Fig. 11). For smaller pump powers ($f = 5...7.5$) more complex two-platicon structures were generated. As for the GVD coefficients, platicon excitation was found to be possible for the rather wide range of these parameters: for $\Delta = 2\kappa_2$, $f = 15$, $|b_{21}| = 0.05$ it was observed for at least $0.3 < |b_{22}/b_{21}| < 5.0$. We also observed a transition from platicons to bright solitons upon frequency scan for $|b_{22}/b_{21}| \ll 1$.

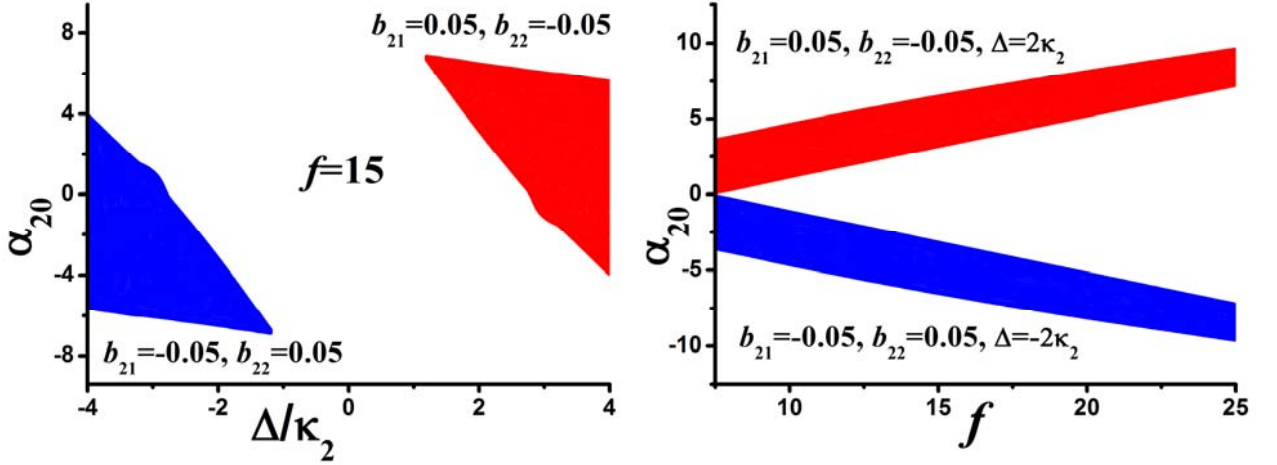

**Fig. 11.** Platicon generation domains for different values of the pumped mode shift and fixed pump amplitude (left panel) and different pump amplitudes and fixed pumped mode shifts (right panel) at $b_{21} = 0.05$, $b_{22} = -0.05$ and $b_{21} = -0.05$, $b_{22} = 0.05$. All quantities are plotted in dimensionless units.

We revealed that this method is less sensitive to the matching of FSRs than the previous one since platicon generation was observed for up to $|d| \sim 2$ at $f = 15$, $\Delta = 2\kappa_2$, $b_{21} = 0.05$, $b_{22} = -0.05$. In contrast with the previous method, in that case the platicon repetition rate may differ from FSRs at the fundamental wave and the second harmonic due to the nonlinear effects. However, their profiles also became asymmetric (see Fig. 12). Also, there was a finite range of the resonant frequency mismatch $\delta$ providing platicon generation. At $f = 15$ and $\Delta = 2\kappa_2$ platicon generation was found at $-25 < \delta < 6$ for $b_{21} = 0.05$, $b_{22} = -0.05$ and at $-6 < \delta < 25$ for $b_{21} = -0.05$, $b_{22} = 0.05$.

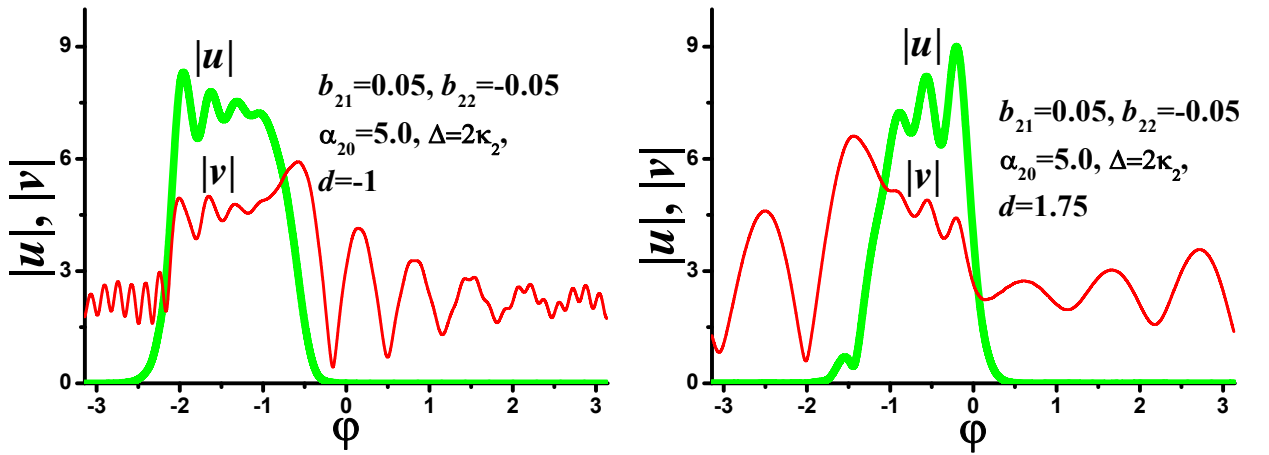

**Fig. 12.** Profiles of the platicon components for different FSRs mismatch values at $f = 15$, $\Delta = 2\kappa_2$, $b_{21} = 0.05$, $b_{22} = -0.05$. All quantities are plotted in dimensionless units.

# CONCLUSION

We demonstrated numerically that the generation of two-color platicons is possible in microresonator-based optical parametric oscillators via pump modulation or mode interaction. The opposite signs of the group velocity dispersion coefficients at interacting harmonics were revealed as a necessary condition for platicon excitation. In contrast with platicons at the SHG process, here platicon generation was observed upon frequency scan only at positive pump detunings for the normal GVD at pump frequency and only at negative detunings in the opposite case. We also found the effect of the transformation of the flat-top profile of the generated platicon component into bell-shaped. It was revealed, that for the efficient platicon excitation one needs simultaneous accurate matching of both microresonator free spectral ranges at interacting harmonics and resonant eigenfrequencies. Excitation conditions and platicon generation domains were found for different generation methods, and properties of generated platicons were studied for different combinations of medium parameters.

# FUNDING

This work was supported by the Russian Science Foundation (Project No. 17-12-01413-П).

# ACKNOWLEDGMENT

The author acknowledges personal support from the Foundation for the Advancement of Theoretical Physics and Mathematics "BASIS".